\documentclass[aps,prl,twocolumn,showpacs,superscriptaddress,preprintnumbers,byrevtex]{revtex4}
\usepackage{xspace}
\usepackage{graphicx}
\usepackage{subfigure}
\usepackage{amsmath}
\usepackage{color}
\def\bz{{B^0}}
\def\taubz{{\tau_\bz}}
\newcommand{\dm}{\Delta m_d}
\newcommand{\dmd}{\dm}
\begin{document}

\preprint{\vbox{ \hbox{   }
    \hbox{Belle Preprint 2008-17}
    \hbox{KEK Preprint 2008-12}
}}

\title{\quad\\[0.5cm] Time-dependent $CP$ Asymmetries in 
$B^0\to K^0_S\rho^0\gamma$ Decays }
\affiliation{Budker Institute of Nuclear Physics, Novosibirsk}
\affiliation{Chiba University, Chiba}
\affiliation{University of Cincinnati, Cincinnati, Ohio 45221}
\affiliation{The Graduate University for Advanced Studies, Hayama}
\affiliation{Hanyang University, Seoul}
\affiliation{University of Hawaii, Honolulu, Hawaii 96822}
\affiliation{High Energy Accelerator Research Organization (KEK), Tsukuba}
\affiliation{Hiroshima Institute of Technology, Hiroshima}
\affiliation{Institute of High Energy Physics, Chinese Academy of Sciences, Beijing}
\affiliation{Institute of High Energy Physics, Vienna}
\affiliation{Institute of High Energy Physics, Protvino}
\affiliation{Institute for Theoretical and Experimental Physics, Moscow}
\affiliation{J. Stefan Institute, Ljubljana}
\affiliation{Kanagawa University, Yokohama}
\affiliation{Korea University, Seoul}
\affiliation{Kyungpook National University, Taegu}
\affiliation{\'Ecole Polytechnique F\'ed\'erale de Lausanne (EPFL), Lausanne}
\affiliation{Faculty of Mathematics and Physics, University of Ljubljana, Ljubljana}
\affiliation{University of Maribor, Maribor}
\affiliation{University of Melbourne, School of Physics, Victoria 3010}
\affiliation{Nagoya University, Nagoya}
\affiliation{Nara Women's University, Nara}
\affiliation{National Central University, Chung-li}
\affiliation{National United University, Miao Li}
\affiliation{Department of Physics, National Taiwan University, Taipei}
\affiliation{H. Niewodniczanski Institute of Nuclear Physics, Krakow}
\affiliation{Nippon Dental University, Niigata}
\affiliation{Niigata University, Niigata}
\affiliation{University of Nova Gorica, Nova Gorica}
\affiliation{Osaka City University, Osaka}
\affiliation{Osaka University, Osaka}
\affiliation{Panjab University, Chandigarh}
\affiliation{Saga University, Saga}
\affiliation{University of Science and Technology of China, Hefei}
\affiliation{Shinshu University, Nagano}
\affiliation{Sungkyunkwan University, Suwon}
\affiliation{University of Sydney, Sydney, New South Wales}
\affiliation{Tata Institute of Fundamental Research, Mumbai}
\affiliation{Toho University, Funabashi}
\affiliation{Tohoku Gakuin University, Tagajo}
\affiliation{Tohoku University, Sendai}
\affiliation{Department of Physics, University of Tokyo, Tokyo}
\affiliation{Tokyo Metropolitan University, Tokyo}
\affiliation{Tokyo University of Agriculture and Technology, Tokyo}
\affiliation{Virginia Polytechnic Institute and State University, Blacksburg, Virginia 24061}
\affiliation{Yonsei University, Seoul}
 \author{J.~Li}\affiliation{University of Hawaii, Honolulu, Hawaii 96822} 
   \author{I.~Adachi}\affiliation{High Energy Accelerator Research Organization (KEK), Tsukuba} 
   \author{K.~Arinstein}\affiliation{Budker Institute of Nuclear Physics, Novosibirsk} 
   \author{T.~Aushev}\affiliation{\'Ecole Polytechnique F\'ed\'erale de Lausanne (EPFL), Lausanne}\affiliation{Institute for Theoretical and Experimental Physics, Moscow} 
   \author{A.~M.~Bakich}\affiliation{University of Sydney, Sydney, New South Wales} 
   \author{V.~Balagura}\affiliation{Institute for Theoretical and Experimental Physics, Moscow} 
   \author{I.~Bedny}\affiliation{Budker Institute of Nuclear Physics, Novosibirsk} 
   \author{V.~Bhardwaj}\affiliation{Panjab University, Chandigarh} 
   \author{U.~Bitenc}\affiliation{J. Stefan Institute, Ljubljana} 
   \author{A.~Bozek}\affiliation{H. Niewodniczanski Institute of Nuclear Physics, Krakow} 
% ...
\collaboration{The Belle Collaboration}

\begin{abstract}
  We report the first measurement of $CP$-violation parameters in 
$B^0\to K_S^0\rho^0\gamma$ decays based on 657 million $B\overline B$ pairs
collected with the Belle detector at the KEKB asymmetric-energy collider.
We measure the time-dependent
$CP$ violating parameter $\mathcal{S}_{K_S^0\rho^0\gamma}=
0.11\pm 0.33(\mathrm{stat.})^{+0.05}_{-0.09}(\mathrm{syst.}) $.
We also obtain the effective direct $CP$ violating parameter 
$\mathcal{A}_\mathrm{eff}=0.05\pm 0.18(\mathrm{stat.})
\pm 0.06(\mathrm{syst.}) $ for 
$m_{K_S\pi^+\pi^-}<1.8$ GeV/$c^2$ and
$0.6\,\mathrm{GeV}/c^2<m_{\pi^+\pi^-}<0.9\,\mathrm{GeV}/c^2$.
\end{abstract}

\maketitle

In the standard model (SM), a mostly left (right)-handed 
photon emitted from a
$\overline B^0$ ($B^0$) meson is expected in the
$b\to s\gamma$ transition.
Hence a small time-dependent
$CP$ asymmetry is predicted in decays of the type
$B\to f_{CP}\gamma$~\cite{Atwood:1997zr}, where
$f_{CP}$ is a $CP$ eigenstate.
New Physics (NP) 
may lead to deviations from the SM expectation by introducing
different photon polarizations in the transition, and can be probed via
experimental measurements of $CP$ violation parameters~\cite{Gershon:2006mt}.
In multi-body final states $B^0\to P^0Q^0\gamma$
with $P^0$ and $Q^0$ being $C$ eigenstates, the same argument
holds~\cite{Atwood:2004jj}.
Measurements of the time-dependent $CP$ asymmetry in $B^0\to K_S^0\pi^0\gamma$
have been reported by Belle and BaBar based on 535 and 431 million
$B\bar B$ pairs~\cite{Ushiroda:2006fi,:2007qj}, respectively.
In this paper, based on 657 million $B\bar B$
pairs collected with the Belle detector~\cite{:2000cg}
at the KEKB asymmetric-energy $e^+e^-$ collider~\cite{KEKB},
we report the measurement of $CP$-violation parameters
on a new channel $B^0\to K_S^0\rho^0\gamma$
where the $B^0$ decay vertex
can be reconstructed from two charged pions from the $\rho^0$ decays.

 At the KEKB, the $\Upsilon(4S)$ is produced with a Lorentz boost of
$\beta\gamma=0.425$ along the $z$ axis, which is defined as the direction
antiparallel to the $e^+$ beam direction.
In the decay chain $\Upsilon(4S)\to B^0\overline B^0\to
 f_\mathrm{sig}f_\mathrm{tag}$, where one of the $B$ mesons decays
at time $t_\mathrm{sig} $ to
the signal mode $f_\mathrm{sig}$ and the other decays at time
$t_\mathrm{tag}$ to a final state $f_\mathrm{tag}$ that distinguishes
$B^0$ and $\overline B^0$, the time-dependent decay rate is given by:
\begin{multline}
\label{eqn:tcpv}
  P(\Delta t,q)=
       \frac{e^{-|\Delta{t}|/{\taubz}}}{4{\taubz}}
\biggl\{1 + q\cdot
\Bigl[ \mathcal{S}\sin(\dmd\Delta{t})\\
   + \mathcal{A}\cos(\dmd\Delta{t})
\Bigr]
\biggr\}.
\end{multline}
Here the probability density function (PDF)
is normalized as a function of two variables: 
$B$ flavor $q=+1(-1)$ when the tagging $B$ meson is $B^0(\overline B^0)$,
and decay time difference $\Delta t$ between two $B$ mesons.
In Eq.(\ref{eqn:tcpv}),
$\mathcal{S}$ and $\mathcal{A}$ 
are $CP$-violation parameters, $\tau_{B^0}$ is the $B^0$ lifetime,
$\Delta m_d$ is the mass difference between the two $B^0$ mass
eigenstates. 
Since the $B^0 \overline B^0$ mesons are approximately at rest in the
$\Upsilon(4S)$ center-of-mass system (c.m.s.), $\Delta t$
can be determined from $\Delta z$, the displacement in $z$ between the
$f_\mathrm{sig}$ and $f_\mathrm{tag}$ decay vertices:
$\Delta t \simeq \Delta z/(\beta\gamma c)$.

The Belle detector~\cite{:2000cg} is a large-solid-angle magnetic spectrometer
that consists of a silicon vertex detector (SVD), a 50-layer central drift
chamber, an array of aerogel threshold Cherenkov counters, and an
electromagnetic calorimeter (ECL) comprised of CsI(Tl) crystals located inside
a superconducting solenoid coil that provides a 1.5 T magnetic field.
An iron flux-return located outside the coil is instrumented to detect
$K_L^0$ mesons and identify muons.

High energy photons are selected from isolated ECL clusters
with no corresponding charged track and c.m.s. energy satisfying
$1.4\,\mathrm{GeV}<E_{\gamma}^*<3.4\,\mathrm{GeV}$.
The photons are also required to lie in the barrel region of the ECL,
and have a photon shower-like shape $E_9/E_{25}>0.95$,
where $E_9$ and $E_{25}$ are the energies summed in $3\times3$ and
$5\times5$ arrays of crystals around the center of the shower, respectively.
To reduce background from $\pi^0\to\gamma\gamma$ or
$\eta\to\gamma\gamma$,  a $\pi^0 (\eta)$ veto is applied with
$\mathcal{L}_{\pi^0}<0.25\,(\mathcal{L}_{\eta}<0.2)$,
where $\mathcal{L}_{\pi^0(\eta)}$
is a $\pi^0(\eta)$ likelihood described 
in Ref.~\cite{Koppenburg:2004fz}.

Neutral kaons ($K_S^0$) are reconstructed from two oppositely charged
pions whose invariant mass lies within 15 $\mathrm{MeV}/c^2$ of the $K_S^0$
nominal mass.  Requirements on impact parameter and vertex displacement
are applied~\cite{Chen:2006nk}.


Charged tracks are required to originate from the vicinity of the
interaction point (IP).  Charged pions
should have kaon and pion identification
likelihoods consistent with the pion hypothesis.
This requirement has an efficiency of $85\%$ with a $7\%$
kaon fake rate.

The decay $B^+\to K^+\pi^-\pi^+\gamma$ is also reconstructed to
study the $K\pi\pi$ system and to serve as a control sample.
Charged kaons are selected from charged tracks and required
to be identified as a kaon with $86\%$ efficiency and an $8\%$ pion fake rate.

We form two kinematic variables: the energy difference
$\Delta E= E_B^*-E_\mathrm{beam}^*$ and the beam-energy
constrained mass $M_\mathrm{bc}=\sqrt{(E^*_\mathrm{beam})^2- (p_B^*)^2}$,
where $E_\mathrm{beam}^*$ is the beam energy in the c.m.s.,
and $E_B^*$ and $p_B^*$ are the c.m.s. energy and momentum of
the reconstructed $B$ candidate, respectively.
The requirement $-0.1\,\mathrm{GeV}<\Delta E<0.08\,\mathrm{GeV}$
is applied.  The $K^+\pi^-\pi^+$ and $K_S^0\pi^-\pi^+$
invariant masses are required to be less than 1.8 GeV/$c^2$.

The $B^0\to K_S^0\rho^0 \gamma$ candidates are selected 
from the $K_S^0\pi^+\pi^- \gamma$ sample by requiring the
$\pi^+\pi^-$ invariant mass to lie in the $\rho^0$ region,
$0.6\,\mathrm{GeV}/c^2<m_{\pi\pi}<0.9\,\mathrm{GeV}/c^2$.
Since the $\rho^0$ is wide,
other modes, such as $K^{*+}\pi^- \gamma$ may also contribute.
We first measure the effective $CP$-violating
parameters, $\mathcal{S}_{\rm{eff}}$ and
$\mathcal{A}_{\rm{eff}}$, using the final sample
and then convert them to the $CP$-violating parameters
of $B^0\to K_S^0 \rho^0\gamma$ using a dilution factor $\mathcal{D}$,
which is discussed later.

In order to suppress the background from light
quark pair production $q\bar q$ ($e^+e^-\to q\bar q$ with
$q=u,d,s,c$), the selection based on an event likelihood ratio
$\mathcal{R}\equiv \mathcal{L}_\mathrm{sig}/
(\mathcal{L}_\mathrm{sig} + \mathcal{L}_\mathrm{bkg})$ is applied.
The likelihood for signal ($\mathcal{L}_\mathrm{sig}$)
and background ($\mathcal{L}_\mathrm{bkg}$) is formed by combining a Fisher
discriminant $\mathcal{F}$ that uses extended modified
Fox-Wolfram moments~\cite{Abe:2003yy}
and the polar angle of the $B$ meson in the
c.m.s. ($\cos\theta_B$).  If there are multiple candidates in
an event, we choose the candidate that has the largest $\mathcal{R}$.
We define $\Delta E$ sideband events with the criteria
$0.1\,\mathrm{GeV}<\Delta E <0.5\,\mathrm{GeV}$
and by vetoing $B\to (K\pi^{\pm})\gamma$.
Other backgrounds that pass our selection criteria are classified
as (a)  $B\to K^*\gamma$ background, (b) $B \to X_s\gamma$ background,
(c) $B\overline B$ background, which includes a generic $b\to c$ component
and charmless $B$-decay component.

The self-cross-feed (SCF) component consists of events in which
not all tracks are from the signal side, in contrast to a true
signal with all tracks correctly assigned.
We define the total signal yield
as the sum of SCF and true signal events.
The fraction of SCF in the total signal yield
ranges from $5.5\%$ to $10.8\% $ in the
$M_\mathrm{bc}$ signal region defined as
$5.27\,\mathrm{GeV}/c^2<M_\mathrm{bc}<5.29\,\mathrm{GeV}/c^2$,
depending on the kaonic resonance.

We obtain the flavor $q$, tagging quality factor $r\in [0,1]$,
and signal and tag side vertices
from the procedure described in Ref.~\cite{:2007jf},
using the two charged pions to reconstruct the signal vertex position.
There is no flavor discrimination when $r=0$, while
the flavor tagging is unambiguous when $r=1$.
Events are sorted into seven $r$-bins.

\begin{figure}
\includegraphics[width=0.7\columnwidth]
{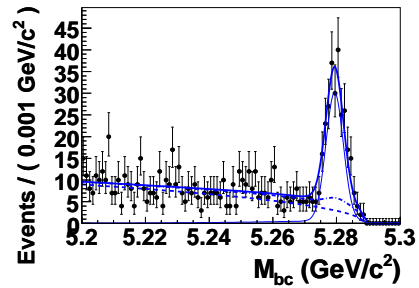}
\caption{\label{fig:ks_vtx-data-mbcs}
 $M_\mathrm{bc}$ distributions for 
 $B^0\to K_S^0\pi^+\pi^-\gamma$ events.
Points with error bars are data.  The curves show the results
from the $r$ dependent $M_\mathrm{bc}$ fit.
The dashed and dash-dotted curves are the $q\bar q$ and all BG.
The thin curve is the total signal including SCF and the thick curve
is the total PDF.
}
\end{figure}

We obtain 299 events in the signal $M_\mathrm{bc}$ region after vertexing.
An unbinned maximum likelihood fit to $M_\mathrm{bc}$ is applied
with signal and $q\bar q$ yields
floated in each $r$-bin.
In the signal $M_\mathrm{bc}$ region,
we find $212\pm 17$ total signal with a fraction of 6.0\% SCF,
$53.4\pm 2.6$ continuum, along with 7.8 $K^*\gamma$,
21.5 other $X_s\gamma$, and 9.0 $B\bar B$ events. 
The true signal shape is parameterized as a Crystal Ball function
with a width corrected using the $B^+\to K^+\pi^-\pi^+\gamma$ data.
The $q\bar q$ shape 
is obtained from the $\Delta E$ sideband events.
For other backgrounds and SCF, shapes and $r$-dependent fractions are
obtained from MC since their contributions are limited.
Figure~\ref{fig:ks_vtx-data-mbcs}
shows the $M_\mathrm{bc}$ projection of the fit.

The $\mathcal{S}_\mathrm{eff}$ and $\mathcal{A}_\mathrm{eff}$ parameters
are extracted from an unbinned maximum likelihood fit
to the $\Delta t$ distribution.  The likelihood function for each event is:
\begin{equation}
\begin{split}
P_i &= (1-f_\mathrm{ol})\int_{-\infty}^{+\infty}d(\Delta t')
\left[\sum_j f_j P_j(\Delta t')R_j(\Delta t_i - \Delta t')\right]\\
& + f_\mathrm{ol}P_\mathrm{ol}(\Delta t_i),
\end{split}
\end{equation}
where $j$ runs over the signal, SCF and four BG components 
($q\bar q$, $K^*\gamma$, other $X_s\gamma$, $B \bar B$).
The fraction of each component ($f_j$) is calculated using
the $r$ dependent $M_\mathrm{bc}$ fit result on an event-by-event basis.
$R_j$ is the resolution function
in the $\Delta t$. 

The PDF for the signal $\Delta t$ distribution $P_\mathrm{sig}$,
is given by a modified form of Eq.(\ref{eqn:tcpv}),
which incorporates the effect
of incorrect flavor assignment.  The parameterization of $R$ is
the same as the one used in the $B^0\to \phi K^0$ analysis~\cite{Chen:2006nk}.
The PDF for SCF ($P_\mathrm{SCF}$) is the same as for the signal
except for using a shorter lifetime $1.16\pm 0.02$ ps,
determined from the MC study.
The same functional forms for the PDF and resolution are used for
the $K^*\gamma$, other $X_s\gamma$ and $B\bar B$ components but
with other lifetime values obtained from MC and $CP$ parameters fixed to zero.
The PDF for $q\bar q$ background events, $P_{q\bar q}$, is modeled
as a $\delta$ function convolved with a double-Gaussian resolution
function $R_{q\bar q}$.  The parameters in $R_{q\bar q}$ is determined
from a fit to $\Delta E$ sideband events.
$P_\mathrm{ol}$ is a Gaussian function that represents a small
outlier component with a fraction $f_\mathrm{ol}$.

\begin{figure}
\includegraphics[width=0.5\columnwidth]
{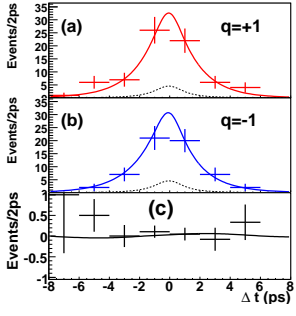}
\caption{\label{fig:data-dt}
 Fit projections on the $\Delta t$ distributions with (a) $q = +1$
and (b) $q = -1$ for events with $r>0.5$. The solid curves are the
fit while the dashed curves show the background contributions.
The raw asymmetry as a function of $\Delta t$ is shown in (c)
with a fit curve superimposed.}
\end{figure}

The only free parameters in the $CP$ fit to $B^0\to K_S^0\pi^+\pi^-\gamma$
are $\mathcal{S}_\mathrm{eff}$ and $\mathcal{A}_\mathrm{eff}$,
which are determined by maximizing the likelihood
function $L= \prod _i P_i$ for events in the $M_\mathrm{bc}$ signal region.
We obtain
$\mathcal{S}_\mathrm{eff} = 0.09\pm 0.27(\mathrm{stat.})
 ^{+0.04}_{-0.07}(\mathrm{syst.})$ and
$\mathcal{A}_\mathrm{eff} = 0.05\pm 0.18(\mathrm{stat.})
 \pm 0.06(\mathrm{syst.})$.
We define the raw asymmetry in each $\Delta t$ bin by
$(N_+ - N_-)/(N_+ + N_-)$, where $N_+$ ($N_-$) is the number of
candidate events with $q= +1(-1)$.
Figure~\ref{fig:data-dt} shows the $\Delta t$ distributions
and the raw asymmetry for events with $0.5< r \leq 1.0$.

Various validity checks for our fitting procedure are performed.
Lifetime fit results for the $B^+\to K^+\pi^-\pi^+\gamma$
and $B^0\to K_S^0\pi^+\pi^-\gamma$ modes are consistent
with the nominal $B^+$ and $B^0$ lifetimes.
No $CP$ asymmetry is seen for the control sample
$B^+\to K^+\pi^-\pi^+\gamma$.

We evaluate systematic uncertainties from the following sources.
The largest contribution is due to the vertex reconstruction, where
the selection criteria are varied to calculate the systematics as
$\Delta \mathcal{S}_\mathrm{eff}=^{+0.01}_{-0.06}$,
$\Delta \mathcal{A}_\mathrm{eff}=\pm 0.03$.
The $\mathcal{S}$
values for $b\to s$ $CP$ modes are varied from 0 to $\sin 2\phi_1$ to
estimate the effect of $CP$ asymmetry in the $B\overline B$ background.
We fit the data with each fixed
parameter shifted by its error to evaluate the uncertainties due to:
background and SCF fractions, shapes, $\Delta t$
PDFs, resolution function, flavor tagging and physics parameters
$\tau_{B^0}$, $\Delta m_d$.  Effects of tag side interference are evaluated
in the same way as in Ref.~\cite{Ushiroda:2006fi}.


The parameter $\mathcal{S}_\mathrm{eff}$
is related to $\mathcal{S}$ for $K_S^0\rho^0\gamma$
with a dilution factor $\mathcal{D}$
that depends on the $K^{*\pm}\pi^{\mp}$ components:
\begin{multline}
\label{eqn:s_d}
\mathcal{D} = \frac{\mathcal{S}_\mathrm{eff}}
{\mathcal{S}_{K_S^0\rho^0\gamma}} =\\
 \frac{\int
[|F_A|^2+ 2\Re(F_A^*F_B) + F_B^*(\bar K)F_B(K) ]}
{\int \left[|F_A|^2+ 2\Re(F_A^*F_B) + |F_B|^2\right]},
\end{multline}
where $F_A,F_B$ are photon-helicity averaged amplitudes for
$B^0\to K_S^0\rho^0(\pi^+\pi^-)\gamma$ and $B^0\to K^{*\pm}(K_S^0\pi^\pm)\pi^\mp\gamma$, respectively.
The factors $F_B(\bar K)$, $F_B(K)$ distinguish between $K^{*-}\pi^+\gamma$ and
$K^{*+}\pi^-\gamma$.  The phase space integral is over the $\rho^0$ region.

The charged mode $B^+\to K^+\pi^-\pi^+\gamma$ is first studied
using a combination of various kaonic resonances with spin $\geq 1$
to model the $K\pi\pi$ system.
The amplitude for a kaonic resonance $K_\mathrm{res}$ that decays into
particle $c$ and subresonance $r$,
where $r$ later decays into particles $a$ and $b$, can be modeled
by the product of two Breit-Wigners:
\begin{equation}
M_{abc|r} = BW(K_\mathrm{res})BW(r)F_K(K_\mathrm{res})F_r f_\mathrm{spin},
\end{equation}
where $BW$ is the relativistic Breit-Wigner lineshape:
$BW(K_\mathrm{res})=1/(m_{K\mathrm{res}}^2-m_{abc}^2-
i m_{K\mathrm{res}} \Gamma_{K\mathrm{res}}) $ and 
$BW(r) = 1/(m_r^2 - m_{ab}^2 - im_{r}\Gamma_{ab})$.
The width
$\Gamma_{ab}= \Gamma_r (q/q_0)^{2L+1}\left(m_r\over m_{ab} \right)F_r^2$
is a function of $q$ and $q_0$, the momenta of particle $a$
in the $r$ rest frame with mass $m_{ab}$ and $m_r$, respectively.
$F_K(K_\mathrm{res})$ and $F_r$ are the Blatt-Weisskopf penetration
factors~\cite{blatt} for resonances $K_\mathrm{res}$ and $r$. 
The spin factor $f_\mathrm{spin}$ for the resonances used in the nominal fit is:
$f_\mathrm{spin}=1$ for $K_1^+(1270)\to K\rho^0,K^*\pi$, $K_1^+(1400)\to K^*\pi$ with a $1^+S$ wave;
$f_\mathrm{spin}=\sin\theta$ for $K^*(1680)\to K\rho^0, K^*\pi$ with a $1^-P$ wave;
$f_\mathrm{spin}=\sin\theta$ for $K_2^+(1430)\to K^*\pi, K\rho^0$ with a $2^+D$ wave.
Here $\theta$ is the helicity angle of subresonance $r$.


Kaonic resonances with different spin-parity waves do not interfere if
the decay plane orientation variables are integrated out. 
The total rate is an incoherent sum of contributions
from different spin-parities with a phase space factor.



To determine the $K^+\rho^0\gamma$ component
in $B^+\to K^+\pi^+\pi^-\gamma$, we study events in a 
$K^*$ region defined as $|m_{K^+\pi^-}-0.8961|< 0.075$ GeV/$c^2$, where
most of the $B^+\to K^+\pi^-\pi^+\gamma$ signal lies.
We fit the $m_{K\pi\pi}$,
$m_{\pi^+\pi^-}$, $m_{K^+\pi^-}$
distributions 
for events in the signal $M_\mathrm{bc}$ region.
In these fits, the yields and shapes for backgrounds are obtained from
the corresponding $M_\mathrm{bc}$ fit results and MC.
The fraction and shape for
SCF are associated with each kaonic resonance from MC.
The yields for the $2^+$ $K_2^*(1430)\gamma$ components are always fixed
based on the measured branching fraction 
$\mathcal{B}(B\to K_2^*(1430)\gamma) =
(1.24\pm0.24)\times 10^{-5}$~\cite{Yao:2006px}.


A two-dimensional (2D) $m_{K\pi},m_{\pi\pi}$ fit is performed in the
$K^*$ region. 
Here the SCF is further categorized depending on whether the $K^+,\pi^-$
tracks are correctly reconstructed.
The floating parameters are the rate and phase for 
$K_1(1270)\to K^*\pi$ relative to $K_1(1270)\to K\rho$.
In the fit, the yields of $1^+,1^-$ kaonic resonances are obtained from
a fit to the $m_{K\pi\pi}$ distribution in the $K^*$ region.
The rates for $K^+\rho^0$ relative to $K^{*0}\pi^+$ 
in the $K^*(1680)$ and $K_2^*(1430)$ resonances
are fixed according to the PDG values~\cite{Yao:2006px}, taking
into account phase-space and isospin factors.
The phases for
$K_\mathrm{res}\to K^{*0}\pi^+$ relative to $K_\mathrm{res}\to K^+\rho^0$
in the $K^*(1680)$, $K_2^*(1430)$ resonances are fixed to be
$-0.36$ rad~\cite{Aston:1986jb} and $-30^\circ$~\cite{Daum:1981hb},
respectively. The PDF for the $m_{K\pi\pi}$ distribution in
the $\rho^0$ region is then obtained.

\begin{figure}
\includegraphics[width=0.6\columnwidth]
{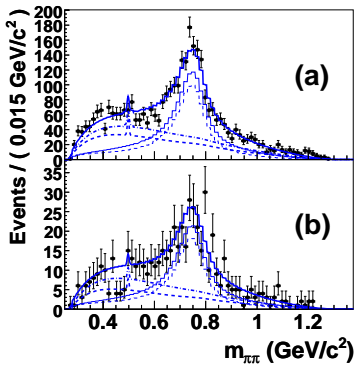}
\caption{\label{fig:data-mpipi}
$m_{\pi\pi}$ distributions for (a) $B^+\to K^+\pi^-\pi^+\gamma$ and
(b) $B^0\to K_S^0\pi^+\pi^-\gamma$ .
The curves follow the convention in Fig.~\ref{fig:ks_vtx-data-mbcs}.
The thin dashed curve is the correctly reconstructed $B\to K_1(1270)\gamma$ signal.
}
\end{figure}

The procedure is repeated for different fixed values of the
phase $\phi$ for $K_1(1400)\to K^*\pi$ relative to $K_1(1270)\to K\rho$.
For the nominal fit, we use $\phi = 15^\circ$ which gives 
the smallest total $\chi^2$ for the 2D $m_{K\pi},m_{\pi\pi}$ distribution
in the $K^*$ region and $m_{K\pi\pi}$ distribution in the $\rho^0$ region.
Figure~\ref{fig:data-mpipi}(a) shows
the distributions for $m_{\pi\pi}$ in
the whole region. The data distributions for $m_{K\pi\pi},m_{K\pi},m_{\pi\pi}$
in other regions also agree with our model.

  Using isospin symmetry, we assume that the fraction and phase
of each kaonic resonance channel in the $B^0$ decay is the same
as that in the $B^+$ decay.
Therefore, using the fit results for the kaonic resonance composition
and substituting the neutral kaonic resonances
and $K_S^0$ mass instead of $K^+$ mass, we obtain
the contributions of the terms $|F_A|^2$, $|F_B|^2$ in Eq.(\ref{eqn:s_d}),
as listed in Table~\ref{tab:ks-data-term-rho0}.
Figure~\ref{fig:data-mpipi}(b) shows the
$m_{\pi\pi}$ distributions for $B^0\to K_S^0\pi^+\pi^-\gamma$.

\begin{table}
\caption{\label{tab:ks-data-term-rho0}
The number of events for each kaonic resonance and
various interference terms 
for the final state $K_S^0\pi^+\pi^-\gamma$.
Interf.\ denotes the interference between 
$K_S^0\rho^0\gamma$ and $K^{*+}\pi^-\gamma$ = $2\Re(F_A^*F_B)$.}
\begin{ruledtabular}
\begin{tabular}{lccccc}
  & Total &$K_S^0\rho^0\gamma$ & $K^{*+}\pi^-\gamma$ & Interf. &
$\begin{array}{c}F_B^*(\bar K)\cdot \\ F_B(K)\end{array}$ \\
\hline
$K_\mathrm{res}(1^+)\gamma$ & 193.6 & 151.0 & 35.1 & 7.5 & 4.4\\
$\;(K_1^0(1270)\gamma)$     & (167.6) & (151.0) & (38.0) & ($-$21.4)
 & (5.2)\\
$K_\mathrm{res}(1^-)\gamma$ & 24.2  & 11.3 & 8.0 & 4.9 & 1.3\\
$K_2^{*0}(1430)\gamma$      & 10.4  & 2.2 &6.1 & 2.0 & 4.5  \\
Sum                         & 228.1 & 164.4 & 49.2 & 14.5 & 10.2
\end{tabular}
\end{ruledtabular}
\end{table}

Various systematic uncertainties in the dilution factor have been investigated.
The dominant one comes from the modeling systematics.  We
put additional resonances to the default model and repeated the
fit procedure.  For $B\to K\sigma\gamma$ in which the $\sigma$ has
a $C$ parity opposite to that of the $\rho^0$, the dilution factor is:
\begin{equation}
\label{eqn:s_d-sys_m}
 \mathcal{D}   = \frac{
\int[ F_{-}(K)^*F_{-}(\bar K)- F_{+}^*F_{+} ]
}
{\int |F_- + F_+|^2} .
\end{equation}
Here $F_{+}$ is the photon-helicity averaged amplitude for
$K_1(1270)(\to K_S^0\sigma)\gamma$, while $F_{-}= F_A+F_B$
[Eq.(\ref{eqn:s_d})].
In the absence of $F_{+}$, Eq.(\ref{eqn:s_d-sys_m})
reduces to Eq.(\ref{eqn:s_d}).
The $\mathcal{D}$ value in this case is calculated to be 1.00.
Other systematics considered include: model uncertainty from 
$K^{*0}(1430)\pi\gamma$, $K_1(1400)\gamma$, $K_1(1270)\to K\omega$ components,
uncertainty in subresonance rates and phases, SCF uncertainty,
isospin breaking, the mass and width
of kaonic resonances, and statistical uncertainty from data.
Adding all the uncertainties in quadrature,
we obtain $\mathcal{D}=0.83^{+0.19}_{-0.03}$.

  By combining $\mathcal{S}_\mathrm{eff}$
and the dilution factor $\mathcal{D}$, we obtain 
$\mathcal{S}_{K_S\rho^0\gamma} = 0.11\pm 0.33(\mathrm{stat.})
^{+0.05}_{-0.09}(\mathrm{syst.}) $.

  In summary, we have measured the time-dependent
$CP$ asymmetry in the decay $B^0\to K_S\rho^0\gamma$ using events with
$m_{K\pi\pi}<1.8$ GeV/$c^2$ and
$0.6\,\mathrm{GeV}/c^2<m_{\pi\pi}<0.9\,\mathrm{GeV}/c^2$. We obtain
$CP$-violation parameters $\mathcal{S}_{K_s\rho^0\gamma}
=0.11\pm 0.33(\mathrm{stat.})^{+0.05}_{-0.09}(\mathrm{syst.}) $
and $\mathcal{A}_\mathrm{eff}=0.05\pm 0.18(\mathrm{stat.})
\pm 0.06(\mathrm{syst.})$.
With the present statistics, the result is consistent with zero and
comparable in precision to the measurement in $B^0\to K_S^0\pi^0\gamma$.
This is the first measurement of $CP$ asymmetry parameters
in the $K_S\rho^0\gamma$ mode
and constrains right-handed currents from NP.

We thank the KEKB group for excellent operation of the
accelerator, the KEK cryogenics group for efficient solenoid
operations, and the KEK computer group and
the NII for valuable computing and SINET3 network
support.  We acknowledge support from MEXT and JSPS (Japan);
ARC and DEST (Australia); NSFC (China); 
DST (India); MOEHRD, KOSEF and KRF (Korea); 
KBN (Poland); MES and RFAAE (Russia); ARRS (Slovenia); SNSF (Switzerland); 
NSC and MOE (Taiwan); and DOE (USA).

\end{document}